%% file: paper.tex
\documentclass{article}

\usepackage{a4wide}
\usepackage{times}
\usepackage{graphics}
\usepackage{verbatim}
\usepackage{alltt}
\usepackage{moreverb}
\usepackage{url}
\usepackage{xspace}
\usepackage{boxedminipage}
\usepackage{lhs2TeXPreamble}
\usepackage{color}

\newtheorem{defi}{Definition}
\newcommand{\dimit}[1]{\textsf{#1}}
\newcommand{\noskip}{\topsep0pt \parskip0pt \partopsep0pt}
\newcommand{\adhoc}{$\mathit{adhoc}$}
\newcommand{\hfoldr}{$\mathit{hfoldr}$}
\newcommand{\strafunski}{\textit{\textsf{Strafunski}}}
\newcommand{\hscaption}[1]{#1\vspace{-22\in}}
\newcommand{\cod}{\ensuremath{c}}

\renewcommand{\qquad}{\hspace{40\in}}

\newcommand{\sforall}{\overline{\forall}}
\newcommand{\defone}{[i]}
\newcommand{\deftwo}{[ii]}

\begin{document}

\title{Strategic polymorphism\\requires just two combinators\emph{!}}

\author{Ralf L{\"a}mmel\,$^{1,2}$ and Joost Visser\,$^{1,3}$\medskip\\
{\small $^1$\ CWI, Kruislaan 413, NL-1098 SJ Amsterdam}\\
{\small $^2$\ Vrije Universiteit, De Boelelaan 1081a, NL-1081 HV Amsterdam}\\
{\small $^3$\ Software Improvement Group, Kruislaan 419, NL-1098 SJ Amsterdam}\\
{\small Email: \texttt{(Ralf.Laemmel|Joost.Visser)@cwi.nl}}%\\
%{\small WWW: \texttt{http://www.cwi.nl/\~{}(ralf|jvisser)/}}\\
%{\small Phone: \texttt{+31 20 592 (4090|4266)} and Fax: \texttt{+31 20 592 4199}}
}

\date{}

\maketitle

%%%%%%%%%%%%%%%%%%%%%%%%%%%%%%%%%%%%%%%%%%%%%%%%%%%%%%%%%%%%%%%%%%%%%%%%%%%%%%

\begin{abstract}

\noindent
In previous work, we introduced the notion of \emph{functional
strategies}: first-class generic functions that can traverse terms of
any type while mixing uniform and type-specific behaviour. Functional
strategies transpose the notion of term rewriting strategies (with
coverage of traversal) to the functional programming
paradigm. Meanwhile, a number of Haskell-based models and combinator
suites were proposed to support generic programming with functional
strategies.

\medskip

\noindent
In the present paper, we provide a compact and matured reconstruction
of functional strategies. We capture strategic polymorphism by just
two primitive combinators. This is done without commitment to a
specific functional language. We analyse the design space for
implementational models of functional strategies. For completeness, we
also provide an operational reference model for implementing
functional strategies (in Haskell). We demonstrate the generality of
our approach by reconstructing representative fragments of the
\strafunski\ library for functional strategies.
\end{abstract}

%%%%%%%%%%%%%%%%%%%%%%%%%%%%%%%%%%%%%%%%%%%%%%%%%%%%%%%%%%%%%%%%%%%%%%%%%%%%%%

\section{Introduction}

In~\cite{LV02-PADL}, we introduced the notion of \emph{functional
strategies} for which we assume the following matured definition
throughout this paper:
\begin{defi}
Functional strategies are functions that
\begin{enumerate}
\item are generic,
\item can mix uniform and type-specific behaviour,
\item can traverse terms, and
\item are first-class citizens. 
\end{enumerate}
\end{defi}
Functional strategies go beyond parametrically polymorphic functions
because of the abilities to traverse terms and to dispatch to
type-specific behaviour. We call this extra polymorphism simply
`strategic polymorphism'. In the present paper, we will capture
strategic polymorphism by just two fundamental function
combinators. Most of our presentation will avoid commitment to a
specific functional language, but we ultimately define a Haskell-based
reference model.

\medskip

\noindent
Functional strategies were derived from the notion of (typed) term
rewriting
strategies~\cite{CELM96,BKK96,VBT98,Maude99,BKKR01,Laemmel02-TGT}. In
fact, the notion of (traversal) strategies can be amalgamated with
different programming paradigms. We use the term \emph{strategic
programming} for generic programming with strategies in whatever
language.  It is at the heart of strategic programming that traversal
schemes are programmer-definable. The native application domain of
strategic programming is language processing, in particular, the
implementation of functionality for program transformations and
analyses; see~\cite{LV02-PADL,Laemmel02-TGR,LV03-PADL} for a few
typical applications. With strategies, one can operate on large
syntaxes or formats in a scalable and flexible manner. Scalability is
implied by genericity, and flexibility by a combinator style that
enables the definition of appropriate traversal schemes. In fact, the
ability to traverse terms while mixing uniform and type-specific
behaviour is beneficial in almost every non-trivial software
application. This is demonstrated in~\cite{LPJ03}, where an approach
to \emph{general purpose generic programming} is based on original
expressiveness for `strategic polymorphism'.

\medskip

\noindent
Our efforts to amalgamate strategic and functional programming are
scoped by the \strafunski\ project.%
\footnote{\strafunski\ home page:
\url{http://www.cs.vu.nl/Strafunski/}~---~\url{Stra} refers to
strategies, \url{fun} refers to functional programming, and their
harmonious composition is a homage to the music of Igor Stravinsky.}
Distributions of the Haskell-centred \strafunski\ bundle for generic
programming and language processing include generative tool support
that, given the programmer-supplied datatypes, generates the code
needed for strategic programming. In previous work, we came up with
different models of functional
strategies~\cite{LV02-PADL,Laemmel02-SPS}. These models differ
regarding the selection of primitive combinators and their types, but
they all share Haskell as the base language.

\medskip

\noindent
In the present paper, we capture strategic polymorphism by just two
combinators:
{\noskip\begin{itemize}\noskip
\item \adhoc\ for type-based function dispatch, and
\item \hfoldr\ for folding over constructor applications.
\end{itemize}}
\noindent
The combinator couple \adhoc\ and \hfoldr\ was identified
in~\cite{Laemmel02-SPS}, and a very similar couple was employed
in~\cite{LPJ03}.\footnote{There are tiny technical differences.
In~\cite{Laemmel02-SPS}, first-class polymorphism is employed as
opposed to proper rank-2 function types in the present paper and
in~\cite{LPJ03}. Also, in~\cite{LPJ03}, type cast is favoured
as opposed to function dispatch in the present paper and
in~\cite{Laemmel02-SPS}. Furthermore, in~\cite{LPJ03}, a
left-associative fold combinator is used as opposed to the
right-associative combinator in the present paper and
in~\cite{Laemmel02-SPS}.} In the present paper, we use these two
combinators for a very compact and matured reconstruction of
functional strategies.  Our goal here is to avoid an invasive
commitment to Haskell. Also, we want to clearly maintain the link to
strategic programming as it was initiated in the context of term
rewriting. The above two combinators can be mapped to Def.~1 as
follows. The combinator \adhoc\ is crucial for mixing uniform and
type-specific behaviour. It can be considered as a disciplined form of
type case. The combinator \hfoldr\ is the mother of all (one-layer,
i.e., non-recursive) traversal. It folds over the immediate subterms
of a constructor application~---~very much in the style of a list
fold. This turns traversal schemes into programmable entities where
ordinary recursive function definition suffices to complete folding
into recursive traversal.

\medskip

\noindent
The paper is structured as follows. In Sec.~\ref{S:sample}, we
approach to the essential expressiveness for strategic polymorphism
via a motivating example. In Sec.~\ref{S:just}, we define the two key
combinators for strategic polymorphism. In Sec.~\ref{S:models}, we
discuss models of functional strategies. Sec.~\ref{S:just} and
Sec.~\ref{S:models} are largely language-independent. In
Sec.~\ref{S:haskell}, we provide a reference model to extend Haskell
with our two combinators. In Sec.~\ref{S:strafunski}, we demonstrate
the power that results from our language extension by reconstructing
representative parts of \strafunski's library. In Sec.~\ref{S:concl},
the paper is concluded.

%%%%%%%%%%%%%%%%%%%%%%%%%%%%%%%%%%%%%%%%%%%%%%%%%%%%%%%%%%%%%%%%%%%%%%%%%%%%%%

\section{Strategic polymorphism~---~a motivating example}
\label{S:sample}

We choose a simple but challenging example that demonstrates all the
characteristics of functional strategies. We will define a combinator
$\mathit{query}$ with an argument $f$, such that $\mathit{query}~f$
performs a top-down, left-to-right traversal to find a subterm that
can be processed by $f$ so that a value is extracted from the
subterm. A suitable subterm has to meet two criteria. Firstly, its
type must coincide with the domain of $f$. Secondly, $f$ applied to
the subterm should not fail, that is, it should not return
$\mathit{Nothing}$.

\medskip

\noindent
This is the Haskell type of the $\mathit{query}$ combinator:

\input{snip/findType.math}

\noindent
There are three universally quantified type parameters:\footnote{For
the purpose of a homogeneous notation, we always use \emph{explicit}
(top-level) universal quantification in all Haskell type signatures.
Note: the type of $\mathit{query}$ is just a plain rank-1 type, that
is, all type variables are indeed quantified at the top-level. We will
later also employ rank-2 types, and thereby go beyond
Haskell~98. Rank-2 types make \emph{explicit} quantification
mandatory. So we decide to switch to explicit quantification all-over
the place.}
{\noskip\begin{itemize}\noskip
\item $\alpha$ denotes the type of `relevant' subterms from which a value is extracted.
\item $\beta$ denotes the type of the term that is eventually passed
to $\mathit{query}$.
\item $u$ denotes the type of the extracted value.
\end{itemize}}
\noindent
The first argument of $\mathit{query}$ is a function of type
$\mathit{\alpha\to\mathit{Maybe}~u}$ that is meant to interrogate
subterms of type $\alpha$ to extract a value of type $u$. The second
argument is the input term of type $\beta$. The result type of
$\mathit{query}$ is the extracted value $u$ (if any), wrapped in the
partiality monad $\mathit{Maybe}$. There are two class constraints
$\mathit{Term}~\alpha$ and $\mathit{Term}~\beta$ which point out that
$\alpha$ and $\beta$ are place-holders for term types. The Haskell
type class $\mathit{Term}$ hosts expressiveness for `strategic
polymorphism', that is, basically the two combinators \adhoc\ and
\hfoldr\ as we will say later.

\medskip

\noindent
Before we describe the actual definition of $\mathit{query}$, let us
first point out that this sort of function is useful for all kinds of
scenarios in program transformation and analysis:
{\noskip\begin{itemize}\noskip
\item 
  Query the type of an abstraction. For instance, from a Java class
  declaration one may want to retrieve the signature of a method with
  a given name.
\smallskip
\item 
  Query the definition of an abstraction. For instance, from an XML
  DTD, one may want to retrieve the content specification of a
  declared element type.
\smallskip
\item 
  Query the focused entity in the course of refactoring. For example,
  from a Cobol program, one may want to retrieve a focused group of
  data description entries.
\end{itemize}}
\noindent
One can easily think of numerous variations on $\mathit{query}$ that
are equally useful, e.g., implementing bottom-up search, or 
transforming terms rather than querying them, and so on.

\medskip

\noindent
Let's turn to the Haskell definition of the $\mathit{query}$ combinator:

\input{snip/findDef.math}

\noindent
We boxed the two combinators that involve strategic
polymorphism. ($\mathit{adhocMTU}$ is a type-specialised variant of
\adhoc, and $\mathit{oneMTU}$ is defined in terms of \hfoldr\ as we
will show later.) For clarity, their names end on ``\emph{MTU}'' to
remind us of the strategy type at hand: \emph{M}onadic
\emph{T}ype-\emph{U}nifying. Monadic style is in place here because of
the partiality of querying. By `type unification' we mean that
querying returns a value of a specific type, regardless of the type of
the input term. The combinator $\mathit{adhocMTU}$ is used to attempt
an application of $f$ to the input term $x$. The composed function
behaves like the polymorphic function $\mathit{const}\
\mathit{Nothing}$ passed as the first argument to $\mathit{adhocMTU}$
except for the type $\alpha$ handled by the function $f$. If the
attempt to apply $f$ results in a value $u$ (first branch of
\textbf{case}), the search is done, and $u$ is returned. Otherwise
(second branch of \textbf{case}), the combinator $\mathit{oneMTU}$ is
used to call $\mathit{query}$ recursively on the children of $x$,
i.e., its immediate subterms. The $\mathit{oneMTU}$ combinator is
meant to attempt application of its argument to each child in
left-to-right order, and returns the result of the first application
that succeeds. We will later see that $\mathit{oneMTU}$ is just one
example of many `one-layer' traversal combinators~---~they can all be
defined in terms of the fundamental combinator \hfoldr.

\medskip

\noindent
To summarise, the $\mathit{query}$ combinator illustrates all
characteristics of strategies (recall 1.--4.\ in Def.~1):
{\noskip\begin{enumerate}\noskip
\item 
  The aspect that the $\mathit{query}$ combinator is \emph{generic},
  i.e., applicable to \emph{all} term types, is expressed by the
  universally quantified $\beta$ in its type.
\smallskip
\item 
  The aspect that the $\mathit{query}$ combinator \emph{mixes uniform
  and type-specific behaviour} is reflected by the application of the
  combinator $\mathit{adhocMTU}$.
\smallskip
\item 
  The aspect that the $\mathit{query}$ combinator \emph{traverses}
  the input term is reflected by the recursive definition of
  $\mathit{query}$ in terms of the traversal combinator
  $\mathit{oneMTU}$.
\smallskip
\item
  The \emph{first-class} status of functional strategies is
  illustrated by their featuring as combinator arguments (cf.\ the
  first argument of $\mathit{adhocMTU}$ and $\mathit{oneMTU}$).
\end{enumerate}}

\noindent
We refer the reader to~\cite{LV02-PADL,Laemmel02-TGR,LV03-PADL} for
many more examples of strategic polymorphism. The present example has
been carefully chosen to cover all aspects of strategic polymorphism,
and to be representative for strategic programming.

%%%%%%%%%%%%%%%%%%%%%%%%%%%%%%%%%%%%%%%%%%%%%%%%%%%%%%%%%%%%%%%%%%%%%%%%%%%%%%

\section{Just two combinators for strategic polymorphism}
\label{S:just}

In the sequel, we define the two combinators \adhoc\ and \hfoldr\ for
strategic polymorphism. The definition is language-independent. We use
a semi-formal style for the semantics of the combinators, and we use
rank-2 types to assign types to the combinators. The interested reader
is referred to~\cite{Laemmel02-TGT} for a formal definition of typed
rewriting strategies in a basically first-order and many-sorted
term-rewriting setting. The below definition makes heavy use of
higher-orderness and polymorphism. This is the key to capturing
strategic expressiveness in just two dedicated combinators as opposed
to the several combinators in previous work.

%%%%%%%%%%%%%%%%%%%%%%%%%%%%%%%%%%%%%%%%%%%%%%%%%%%%%%%%%%%%%%%%%%%%%%%%%%%%%%

\subsection*{The \adhoc\ combinator for type-based function dispatch}

We want to give a single definition of \adhoc, which is valid for all
different types of functional strategies (think of querying vs.\
transformation).  So we assume the general type scheme $\sforall
\alpha.\ \alpha\to\cod~\alpha$ for functional strategies. Here, $\cod$
is a type constructor that derives the co-domain of the function type
from the domain $\alpha$. If we instantiate the co-domain constructor
$\cod$ with the identity type constructor (i.e., $I~\alpha=\alpha$),
we obtain the type of type-preserving strategies. The constant type
constructor (i.e., $C~u~\alpha=u$) handles the type-unifying scheme
(i.e., the result type is always $u$ regardless of the input type
$\alpha$). We use an over-lined version $\sforall$ of the universal
quantifier $\forall$ to distinguish `strategic polymorphism' from
ordinary parametric polymorphism.

\medskip

\noindent
\ \ \ $\begin{array}{l}
\begin{array}{lcll}
\mathit{adhoc} & :   & \forall{\cod}. & \mbox{{-}{-}\ Co-domain constructor}\\
               &     & \sforall\beta.& \mbox{{-}{-}\ Term type of specific functionality}\\
               &     & (\sforall\alpha.~\alpha\to\cod~\alpha) &
\mbox{{-}{-}\ Generic default}\\
               & \to & (\beta\to\cod~\beta) &
\mbox{{-}{-}\ Type-specific ad-hoc case}\\
               & \to & (\sforall\gamma.~\gamma\to\cod~\gamma)
& \mbox{{-}{-}\ Constructed strategy}
\end{array}\\
\\
\mathit{adhoc}\ p\ m\ x\ =\ \left\{
  \begin{array}{ll}
  m\ x, & \mbox{if}\ \mathit{typeOf}(x)=\mathit{domOf}(m)\\
  p\ x, & \mbox{otherwise}
  \end{array}\right.
\end{array}$\hfill\defone

\medskip

\noindent
The placing of the various $\sforall$ quantifiers in the (rank-2) type
of \adhoc\ emphasises that the combinator constructs a
polymorphic function from a polymorphic function argument $p$ and a
monomorphic function argument $m$.\footnote{Reminder:
$\forall$/$\sforall$ quantifiers expand to the right as far as
possible. Lifting the $\sforall\gamma$ to the top of the function type
would be acceptable, because $\sforall\gamma$ is placed on the right
of the outermost $\to$, and hence, lifting leads to an equivalent
type. By contrast, moving the inner $\sforall\alpha$ to the top would
lead to a too liberal function that also accepted monomorphic
arguments for $p$. Dually, pushing the $\sforall\beta$ to the second
argument (where it is used) would lead to a too restrictive function
that insisted on polymorphic ad-hoc cases for $m$ instead of
type-specific functionality.} In the definition, the expressions
$\mathit{typeOf}(x)$ and $\mathit{domOf}(m)$ denote the
\emph{specific} type of $x$ and the \emph{specific} domain of $m$,
i.e., the respective types to which $\gamma$ and $\beta$ are
instantiated \emph{at run-time}. So, following this definition,
$\mathit{adhoc}~p~m~x$ dispatches to the monomorphic $m$ if $m$ is
applicable to the $x$ at hand, but otherwise it resorts to the
polymorphic, generally applicable $p$.

\medskip

\noindent
In strategic programming, the common usage of \adhoc\ is to derive
\emph{generic} `rewrite steps' from \emph{type-specific}
ones~\cite{LV02-DP-SF}. The mono\-morphic ingredient $m$ is then a
function which rewrites terms of a specific type, potentially based on
pattern matching, whereas the polymorphic ingredient $p$ provides a
trivial generic default. As a simple example, consider the following
polymorphic function $\mathit{negbool}$ which behaves like the
identity function by default but applies Boolean negation ``$\neg$''
when faced with a Boolean:
\[
\begin{array}{l}
\mathit{negbool}\ =\ %
\mathit{adhoc}\;\mathit{id}\;(\neg)
\end{array}
\]
For this type-preserving strategy, the co-domain type constructor
$\cod$ is instantiated to the identity type constructor $I$. In our
$\mathit{query}$ sample, \adhoc\ was used to construct a type-unifying
generic rewrite step
``$\Varid{adhoc}\;(\Varid{const}\;\Conid{Nothing})\;\Varid{f}$''.
Here, the co-domain constructor $\cod$ is instantiated to the constant
type constructor $C$. The generic default
$\Varid{const}\;\Conid{Nothing}$ models the failure of the
$\mathit{query}$. The function $f$ interrogates terms of a certain
type to extract a $\mathit{Maybe}$ value.

%%%%%%%%%%%%%%%%%%%%%%%%%%%%%%%%%%%%%%%%%%%%%%%%%%%%%%%%%%%%%%%%%%%%%%%%%%%%%%

\subsection*{The \hfoldr\ combinator for folding over constructor applications}

Let us recall the basic idiom of list traversal. Without loss of
generality, we consider \emph{right-associative} list
traversal. Folding a list $[x_1,x_2,\ldots,x_n]$ according to
ingredients $f$ and $z$ for the non-empty and the empty list form is
defined as usual:

\[\begin{array}{l}
\mathit{foldr}\ f\ z\ [x_1,x_2,\ldots,x_n]\ =\ f\ x_1\ (f\ x_2\ (\cdots\ (f\ x_n\ z)\ \cdots))
\end{array}\]

\noindent
Folding over constructor applications is similar: instead of folding
over the elements of a homogeneous list, we fold over the children of
a term, i.e., its immediate subterms. Since the children of a term are
potentially of different types, we need a \emph{heterogeneous}
fold. So we use the name \hfoldr. Without loss of generality, we
assume \emph{curried} constructor applications. Folding a term $C\
x_n\ \cdots\ x_2\ x_1$ is now defined as follows:

\medskip

\noindent
\ \ \ $\begin{array}{l}
\begin{array}{lcll}
\mathit{hfoldr} & : & \forall{\cod}.& \mbox{{-}{-}\ Co-domain constructor}\\
                &   & (\sforall\alpha\forall\beta.~\alpha\to\cod~(\alpha\to\beta)\to\cod~\beta)\ \  & \mbox{{-}{-}\ `cons' case}\\
                & \to & (\forall\gamma.~\gamma\to\cod~\gamma) & \mbox{{-}{-}\ `nil' case}\\
                & \to & (\sforall\delta.~\delta\to\cod~\delta)\ \ %
& \mbox{{-}{-}\ Constructed strategy}
\end{array}\\
\\
\mathit{hfoldr}\ f\ z\ (C\ x_n\ \cdots\ x_2\ x_1)\ =\ %
f\ x_1\ (f\ x_2\ (\cdots\ (f\ x_n\ (z\ C))\ \cdots))\\
\mbox{}
\end{array}$\hfill\deftwo

\medskip

\noindent
The indices in $C\ x_n\ \cdots\ x_2\ x_1$ clarify that we treat the
children in the curried expression as a `snoc list' with the rightmost
child as `head'. Also note that the empty constructor application $C$
is passed to $z$ so that the constructor can contribute to the result
of folding. The type of \hfoldr\ is somewhat involved, especially
regarding the argument $f$. Here, $\alpha$ denotes the type of the
current head $x_i$, i.e., the next subterm, and $\beta$ denotes the
type of a fragment of $C~x_n~\cdots~x_i$. The type
$\cod\,(\alpha\to\beta)$ denotes the recursively processed tail, and it
reflects that this tail lacks the $i^\mathit{th}$ child of type
$\alpha$. The type of \hfoldr's argument $z$ of \hfoldr\ is less of a
headache. Since $z$ is meant to process constructors, the $\gamma$ in
its type denotes the type of the constructor $C$, applied to no
children.

\medskip

\noindent
This folding operation is now sufficient to define arbitrary
\emph{one-layer} traversal combinators which in turn can be completed
into recursive traversal schemes in different ways. In the sample
section, we assumed a one-layer traversal combinator $\mathit{oneMTU}$
which can now be defined concisely as follows:
\begin{eqnarray*}
\Varid{oneMTU}\;\Varid{s} & = & \Varid{hfoldr}\;(\lambda \Varid{h}\;\Varid{t}\to \Varid{t}\mathbin{`\Varid{mplus}`}\Varid{s}\;\Varid{h})\;(\Varid{const}\;\Varid{mzero})
\end{eqnarray*}
Here we assume an extended monad with operations $\mathit{mplus}$ for
a kind of choice, and $\mathit{mzero}$ for failure. The
$\mathit{Maybe}$ monad is a typical representative of this class of
monads. The definition states that the argument strategy $s$ is
applied to the head $h$ of the constructor application, and
$\mathit{mplus}$ is used to combine the recursively processed tail $t$
and the processed head. Folding starts from $\mathit{mzero}$.  In
Sec.~\ref{S:strafunski}, we will provide definitions of more one-layer
traversal combinators, and we derive several typical recursive
traversal combinators~---~just in the same way as $\mathit{query}$ was
derive from $\mathit{oneMTU}$ by means of ordinary recursive function
definition.

%%%%%%%%%%%%%%%%%%%%%%%%%%%%%%%%%%%%%%%%%%%%%%%%%%%%%%%%%%%%%%%%%%%%%%%%%%%%%%

\section{Implementational models~---~a detailed analysis}
\label{S:models}

The two combinators \adhoc\ and \hfoldr, which capture strategic
polymorphism, can be modelled in several ways. In this section, we
will analyse the dimensions of this design space for
implementation. This also allow us to refer to the large body of
related work on generic programming. The exploration will avoid
commitment to a specific functional language. This will clarify that
different typed functional languages can be made fit for strategic
programming, e.g., Clean, Haskell, and SML.

%%%%%%%%%%%%%%%%%%%%%%%%%%%%%%%%%%%%%%%%%%%%%%%%%%%%%%%%%%%%%%%%%%%%%%%%%%%%%%

\subsubsection*{Models at a glance}

As with any generic functionality, there are three ways to enable
functional strategies in a given language:
\begin{description}
\item[\dimit{built-in}] The combinators \adhoc\ and \hfoldr\ are
implemented as language primitives.
\item[\dimit{defined}] They are defined in terms of already available
expressiveness.
\item[\dimit{per-type}] They are defined using a term interface which
is implemented per datatype.
\end{description}
Let us make some important side remarks regarding these overall
approaches. The \dimit{built-in} option is the preferred one but it
requires changing the language and its implementations. The
\dimit{defined} option may fail to be faithful or may require
inconvenient encodings, depending on the given expressiveness (as
demonstrated below). To be practical and scalable, the
\dimit{per-type} option needs generative tool support to implement the
term interface per datatype. If the generative component becomes an
integral part of the language implementation (as is the case for other
generic functionality, e.g., equality predicates in SML and
Haskell~98~\cite{HaskellReport}), the \dimit{per-type} option evolves
into the \dimit{built-in} option.

\medskip

\noindent
Whichever option is chosen, another seven dimensions span a design
space:
{\noskip\begin{description}\noskip
\item[type reflection] How to query type information and how to perform coercion?
\item[term reflection] How to generically destruct and construct terms?
\item[application] How to \emph{apply} a strategy to an actual term?
\item[quantification] How to separate strategic and parametric polymorphism?
\item[ranking] How to rank the types of strategy combinators?
\item[reduction] How to deal with eager vs.\ lazy reduction and with effects?
\item[modularisation] How to maintain separate compilation?
\end{description}}

%%%%%%%%%%%%%%%%%%%%%%%%%%%%%%%%%%%%%%%%%%%%%%%%%%%%%%%%%%%%%%%%%%%%%%%%%%%%%%

\subsubsection*{The type-reflection dimension}

Type-based function dispatch (\adhoc) assumes some \emph{type
information} at run-time. In addition, \emph{coercion} of
type-specific cases is needed to apply them to terms encountered at
run-time. Dynamic typing~\cite{ACPP91,ACPR92} provides expressiveness
to \dimit{define} \adhoc. Then, the terms that are processed by
strategies had to be of type \texttt{Dynamic}, and \adhoc\ is
implemented by `dynamic type case' for type matching and
coercion. (There are few language implementations with built-in
support for dynamic typing but Yale Haskell used to support it, and it
was recently added to Clean. \dimit{Per-type} support for dynamic
typing has been suggested in various ways.) In fact, dynamic typing is
more powerful than needed for \adhoc\ because it involves the special
type \texttt{Dynamic}. A lightweight dynamic-typing approach is to
maintain a universe in which all datatypes are embedded~\cite{Yang98}.
This approach is particularly suited for \dimit{per-type}\ generative
support. Embedding can be performed in two ways: either via a
constructor per type~\cite{BKKR01}, or on the basis of a universal
term representation that includes a type
\emph{representation}~\cite{LV02-PADL}. As an alternative to dynamic
typing, one may also consider intensional
polymorphism~\cite{HM95,Weirich02} as a means to \dimit{define}
\adhoc.  This is a major language extension. (It is not available in
widespread language implementations.) More seriously, this approach is
not applicable because all work in the area of intensional type
analysis favours \emph{structural} type analysis while our kind of
dispatching requires \emph{nominal} type analysis as argued
in~\cite{Glew99,LPJ03}. This closes the case to \dimit{define} \adhoc\
in terms of other expressiveness. As for a \dimit{built-in} \adhoc,
the following approaches are at our disposal. Access to run-time type
information can be based on a term representation with type
tags~\cite{Glew99,Laemmel02-TGT}. We should note that such tags are in
conflict with type erasure. Also, they slightly enlarge the run-time
representation of terms, and in turn slow down term
manipulation. Alternatively, term constructors can be used to retrieve
type information via a mapping from constructors to type tags.  Yet
another approach is to rely on carrying dictionaries in the sense of
type classes~\cite{PJ93} instead of carrying types in the terms
themselves. Yet another approach is to rely on run-time type
information complemented by unsafe type coercion as discussed
in~\cite{LPJ03}.

%%%%%%%%%%%%%%%%%%%%%%%%%%%%%%%%%%%%%%%%%%%%%%%%%%%%%%%%%%%%%%%%%%%%%%%%%%%%%%

\subsubsection*{The term-reflection dimension}

The poor man's way to observe term structure is based on a universal
term representation. Here, explosion and implosion functions mediate
between the programmer-supplied datatypes and the universal term
representation. This expressiveness is usually provided
\dimit{per-type}~\cite{BKKR01,LV02-PADL}. Folding over constructor
applications boils then down to ordinary folding over homogeneous
lists of term representations. Note that it is imperative to
effectively hide the representation type in order to guarantee
`implosion safety'~\cite{LV02-PADL}. Then, this approach shines
because of its simplicity.  Note that implosion and explosion are
likely to lead to a performance degradation but the choice of a
suitable representation type can limit the depth of term conversion to
a traversal's extent. A \dimit{built-in} definition of \hfoldr\ would
entirely avoid this rather indirect style of operating on terms. In
fact, built-in support is straightforward: Def.~\deftwo\ can be
defined on any run-time term representation as is. As our
Haskell-specific reference model of the upcoming section will
demonstrate, term reflection can also be supported elegantly
\dimit{per-type}. That is, \hfoldr\ directly operates on terms on the
basis of a per-type implementation of Def.~\deftwo. The most prominent
expressiveness to attempt a \dimit{definition} of \hfoldr\ is
presumably
polytypism~\cite{JJ96,JJ97,Hinze00POPL,Hinze00MPC,HPJ01,MP01} as
implemented in PolyP and Generic Haskell. The combinator \hfoldr\ can
indeed be expressed by structural induction on the type for its
traversed argument. However, nominal run-time type case as needed for
\adhoc\ is beyond the scope of polytypism. The various constructs to
customise polytypic definitions\footnote{Cf.\ ad-hoc definitions for
Generic Haskell as of~\cite{Hinze00POPL}, type-specific instances for
derivable type classes~\cite{HPJ01,MP01} just as for ordinary Haskell
type or constructor classes~\cite{WB89}, copy lines and constructor
cases~\cite{CL02} added to Generic Haskell as of~\cite{Hinze00MPC}.}
are compile-time means as opposed to a combinator for run-time
type-based dispatch. One may also consider recent proposals for
generalised pattern-match constructs~\cite{Jay01,DV01} to define
\hfoldr. Pattern matching is then not restricted to a single type, and
a pattern-match case does not insist on a specific constructor. Again,
these approaches do not offer type case.

%%%%%%%%%%%%%%%%%%%%%%%%%%%%%%%%%%%%%%%%%%%%%%%%%%%%%%%%%%%%%%%%%%%%%%%%%%%%%%

\subsubsection*{The application dimension}

Ideally, strategies are plain functions on the programmer's term
types.  In fact, this characterises a challenging corner in our
multi-dimensional design space. There are the following reasons why
strategy application might deviate from function application. (a)
Strategies might operate on \texttt{Dynamic} or a representation type
in the interest of type and/or term reflection. (b) Strategies might
be wrapped inside datatype constructors for reasons of
opaqueness~\cite{LV02-PADL}, or for reasons of rank-2
polymorphism~\cite{Laemmel02-SPS}. (c) Strategies might operate on
datatypes constructed from term types for reasons of a uniform
definition~\cite{Laemmel02-SPS,LPJ03}: recall the co-domain
constructor \cod\ in Def.~\defone\ and Def.~\deftwo. This constructor
will be normally a proper datatype because type-level lambdas are
hardly supported in functional programming. Strategy application
differs for (a)--(c). As for (a), to deal with \texttt{Dynamic} or a
representation type, the ordinary terms need to be converted before
and after strategy application. For convenience, this can be
encapsulated via an overloaded application
operator~\cite{LV02-PADL}. Then the programmer does not need to
provide type tags. As for (b), a corresponding application operator is
trivially defined by unwrapping. Basic strategy combinators also need
to perform wrapping and unwrapping all-over the place. As for (c), we
are saved by the fact that usually only a small number of co-domain
constructors are used. Hence, one can specialise the types of \adhoc\
and \hfoldr\ for these few cases instead of postponing the type
adjustment until strategy application. The \dimit{built-in} approach
can hide this problem via a closed-world assumption regarding
co-domain constructors.

%%%%%%%%%%%%%%%%%%%%%%%%%%%%%%%%%%%%%%%%%%%%%%%%%%%%%%%%%%%%%%%%%%%%%%%%%%%%%%

\subsubsection*{The quantification dimension}

It is clear that strategies go beyond parametrically polymorphic
functions. To reflect this fact, we used $\sforall$\ instead of
$\forall$ in Def.~\defone\ and Def.~\deftwo. In an actual language
design, we can extend the interpretation of the universal
quantifier, i.e., we equate our $\sforall$ with $\forall$. This is
common practice for intensional polymorphism. We introduced $\sforall$
for the sake of a clear separation of parametric and strategic
polymorphism. Having in mind a \dimit{per-type} approach, we can
also introduce explicit type constraints in the sense of ad-hoc
polymorphism~\cite{WB89}, as supported by Haskell and Clean, i.e., we
equate $\sforall\alpha.~\ldots$ with
$\forall\alpha.\mathit{Term}~\alpha~\Rightarrow~\ldots$. Thus, the
class constraint $\mathit{Term}~\alpha$\ points out where we go beyond
parametric polymorphism. A \dimit{built-in} approach does not rely on
type classes or class constraints.

%%%%%%%%%%%%%%%%%%%%%%%%%%%%%%%%%%%%%%%%%%%%%%%%%%%%%%%%%%%%%%%%%%%%%%%%%%%%%%

\subsubsection*{The ranking dimension}

To enable the combinator style of strategic programming, it is
indispensable that strategy combinators consume polymorphic
arguments. The most basic example is the \hfoldr\ combinator which
must insist on a polymorphic first argument because it is applied to
children of potentially different term types. Normally, the need for
polymorphic function arguments necessitates second-order
polymorphism~\cite{GirardPhD,Reynolds74} as opposed to `simple'
polymorphism with top-level quantification. One can attempt to
organise generic traversal in terms of rank-1
expressiveness~\cite{Jay01} but this will rule out the key idioms of
strategic programming~---~in particular one-layer traversal.
As an aside, second-order polymorphism is generally avoidable if
strategies are weakly typed as functions on a representation
type. Some form of rank-2 types (or even higher ranks) are supported
in several functional language implementations. A well-understood form
is first-class polymorphism~\cite{Jones97} as employed for modelling
functional strategies in~\cite{Laemmel02-SPS}. First-class
polymorphism means to wrap up polymorphic functions as constructor
components. This necessitates unwrapping prior to function
application. In~\cite{LPJ03}, we employ rank-2 types (as supported in
the current GHC implementation of Haskell) for generic traversal
combinators. Regarding the earlier discussion of expressiveness to
\dimit{define} our combinators, we should now add that `second order'
is also indispensable if existing forms of polymorphism were
considered. In particular, polytypism in Generic
Haskell~\cite{Hinze00POPL,Hinze00MPC} is not second order
because polytypic functions cannot involve polytypic function
arguments.

%%%%%%%%%%%%%%%%%%%%%%%%%%%%%%%%%%%%%%%%%%%%%%%%%%%%%%%%%%%%%%%%%%%%%%%%%%%%%%

\subsubsection*{The reduction dimension}

Functional strategy combinators were inspired by term rewriting
strategies~\cite{VBT98,Laemmel02-TGT} which are eager. As for ordinary
function application, the eager functional programmer can effectively
postpone strategy application by the lazy ``$\mathit{if}$'' in the
definition of new strategy combinators. When adding the \hfoldr\
combinator in an eager framework, Def.~\deftwo\ must be implemented
with some care to prevent premature evaluation of applications to
subterms. This is not a problem in a straightforward inductive
implementation (as opposed to the maybe too eager reading of
Def.~\deftwo). There is also no problem with using impure effects such
as in SML rather than monadic effects. This is again demonstrated by
strategic term rewriting \`a la Stratego because Stratego supports
some effects, e.g., for hygenic name generation or I/O.

%%%%%%%%%%%%%%%%%%%%%%%%%%%%%%%%%%%%%%%%%%%%%%%%%%%%%%%%%%%%%%%%%%%%%%%%%%%%%%

\subsubsection*{The modularisation dimension}

Strategic programs do not require compile-time specialisation of
generic functionality as opposed to polytypic programs. This is
because strategic programs operate on the programmer-supplied
datatypes only via \adhoc\ and \hfoldr. These combinators, in turn,
either operate on suitable run-time term representations as
\dimit{built-in}s, or they are overloaded \dimit{per-type}. Hence,
separate compilation is maintainable. Some techniques for type or term
reflection might however imply a closed-world assumption. Firstly, the
definition of a universe with embedding constructors per type is not
extensible unless we assume extensible
datatypes. In~\cite{Laemmel02-SPS}, we deal with the same problem:
type case is encoded in a way that relies on a class member per term
type. Separate compilation is also sacrificed when one provides
\dimit{per-type} functionality by a `monster switch', that is, by a
central authority which would be meant to cover all types as opposed
to proper overloading with support for separate compilation.

%%%%%%%%%%%%%%%%%%%%%%%%%%%%%%%%%%%%%%%%%%%%%%%%%%%%%%%%%%%%%%%%%%%%%%%%%%%%%%

\section{A reference model for Haskell}
\label{S:haskell}

We will now define a Haskell-based reference model for the
implementation of the combinators \adhoc\ and \hfoldr. In view of the
previous section, we can provide the following characterisation:
\begin{itemize}
\item The reference model relies on \dimit{per-type} functionality.
\item Strategies directly operate on the terms of the programmer-supplied datatypes.
\item Strategy application is plain function application.
\end{itemize}
To start with, we define the type scheme for generic functions that
model strategies:

\input{snip/Generic.math}

\noindent
These two type synonyms make a distinction between unconstrained,
i.e., parametrically polymorphic functions and constrained, i.e.,
strategically polymorphic functions. The class constraint points out
where we go beyond parametric polymorphism. Roughly, the
$\mathit{Term}$ class hosts our two combinators, but the details
follow below. Let us first list the Haskell types of our two
combinators:

\input{snip/ComprehensibleTerm.math}

\noindent
The two type synonyms above are defined for convenience to make the
type of \hfoldr\ more comprehensible. In the types of the combinators,
we make use of rank-2 polymorphism as provided by the GHC
implementation of Haskell. It remains to define the combinators, to
provide the complete declaration of the $\mathit{Term}$ class, and to
describe the derivation of the $\mathit{Term}$ instances. We start
with the implementation of Def.~\defone\ for \adhoc. To this end, we
assume an operation $\mathit{typeOf}$ which maps `typeable' values of
any term type to a type representation. Note that this is the kind of
mapping that we proposed earlier in order to avoid carrying type
information in the terms themselves. The operation $\mathit{typeOf}$
is placed in a $\mathit{Typeable}$ class as follows:

\input{snip/Typeable.math}

\noindent
The type $\mathit{TypeRep}$ models type representations. Furthermore,
we assume an operation $\mathit{unsafeCoerce}$ to cast a value of any
type to another type. Then, Def.~\defone\ can be rephrased in Haskell
as follows:

\input{snip/adhocDynamic.math}

\noindent
It is important to notice that the argument of $\mathit{typeOf}$ is
only used to carry type information via overloading. The assumed two
features are readily available in Haskell implementations. Of course,
a proper language extension, which offers \adhoc\ as a
\dimit{built-in}, does not need to expose either $\mathit{typeOf}$ or
$\mathit{unsafeCoerce}$. These two operations are folklore in the
Haskell community because they form the foundation of the
$\mathit{Dynamic}$ library, which has been a standard part of Haskell
distributions for several years. The folk's wisdom to derive
$\mathit{Typeable}$ and to perform type-safe cast on top of it with
the help of $\mathit{unsafeCoerce}$ is found in~\cite{LPJ03}. One can
also use other, maybe safer, but also more involved approaches to
dynamic typing. Regardless of the specific approach, the important
thing to remember is that our implementational model of \adhoc\ only
involves term types but no universe such as \texttt{Dynamic}. This
means that \adhoc\ is very simple in nature, and \adhoc\ is not inherently
dynamically typed.

\medskip

\noindent
From the above development it is clear that \adhoc\ is indeed
\dimit{defined} in terms of \dimit{per-type} functionality for
accessing run-time type representations based on overloading. Then,
the class $\mathit{Term}$ that captures strategic polymorphism has to
be constrained by the $\mathit{Typeable}$ class so that every term
type is also known to be typeable. Thus, we have:

\input{snip/TT.math}

\noindent
Def.~\deftwo\ for the \hfoldr\ combinator immediately necessitates
\dimit{per-type} support because Haskell does not offer any
expressiveness to generically observe the structure of terms. We place
\hfoldr\ itself as a member in the $\mathit{Term}$ class. In fact, the
very comprehensible type of \hfoldr\ from above is not immediately
suited for an overloaded class member. So we place a primed member in
the class with an equivalent type:

\input{snip/Term.math}

\noindent
We define \hfoldr\ simply in terms of $\mathit{hfoldr}'$:

\input{snip/Primed.math}

\noindent
The type of the primed member uses \emph{implicit} quantification over
the class parameter $\alpha$. The non-primed version uses
\emph{explicit} quantification hidden in $\mathit{Strategic}$. The
$\mathit{Term}$ class exhibits an intriguing feature: it is defined
recursively in the sense that the $\mathit{Term}$ class itself is used
to constrain the signature of its member $\mathit{hfoldr}'$ hidden in
$\mathit{HCons}$. This can be viewed as a sign of the first-class
status of strategies.

\medskip

\noindent
The classes $\mathit{Term}$ and $\mathit{Typeable}$ can be now added
to the Haskell~98~\cite{HaskellReport} language definition in the same
way as the standard classes $\mathit{Eq}$, $\mathit{Ord}$,
$\mathit{Show}$, and $\mathit{Read}$. Just as the Haskell language
definition contains a specification of derived instances for these
\dimit{built-in} classes, so do we need to complete our extension by
specifying the derived $\mathit{Term}$ instances. The derivation of
the member $\mathit{hfoldr}'$ is completely straightforward. We need
to provide one equation per constructor based on the scheme for
heterogeneous, right-associative fold according to Def.~\deftwo. That
is, given a constructor $C$ of type $\tau$ with $n$ arguments, we need
a pattern-match case defined as follows:
\[\begin{array}{l}
\mathbf{instance}\ \mathit{Term}\ \tau\ \mathbf{where}\\
\ \ \ \mathit{hfoldr}\ f\ z\ (C\ x_n\ \cdots\ x_2\ x_1)\ =\ f\ x_1\ (f\ x_2\ (\cdots\ (f\ x_n\ (z\ C))\ \cdots))\\
\ \ \ {-}{-}\ \mbox{Continue for the other constructors}
\end{array}\]
This simple scheme is applicable to mutually recursive, parameterised
(perhaps over higher-kinded type variables) datatypes. Not even
non-uniform recursion is a problem. For all basic datatypes such as
$\mathit{Int}$, we assume instances that apply the nil case.  This is
also a sensible choice for function types as there is no way to
traverse them but it should be safe to encounter them in the
course of traversal.

\medskip

\noindent
For completeness, we have investigated the option to \dimit{define}
our combinators by means of \emph{derivable type classes} as they were
proposed for Haskell and Clean~\cite{HPJ01,MP01}. Derivable type
classes are precisely meant for the definition of classes for which
the instances follow a common scheme. To this end, polytypic patterns
for sums and products~\cite{JJ96,JJ97} are included in the
pattern-match syntax for class member definition. The \adhoc\
combinator (and hence, the $\mathit{Typeable}$ class) does not take
advantage of derivable type classes because it necessitates a nominal
approach rather than structural induction. However, the \hfoldr\
combinator (and hence, the $\mathit{Term}$ class) can be defined as
follows:

\input{snip/dtc-hfoldrTP.math}

\noindent
There are equations for $\mathit{Unit}$, $\alpha~{:}{+}{:}~\beta$
(i.e., sums), and $\alpha~{:}{*}{:}~\beta$ (i.e., right-associative
products). In fact, we only show the special case for the
type-preserving scheme where the co-domain constructor is instantiated
to the identity type constructor $I$.

%%%%%%%%%%%%%%%%%%%%%%%%%%%%%%%%%%%%%%%%%%%%%%%%%%%%%%%%%%%%%%%%%%%%%%%%%%%%%%

\section{Reconstruction of \strafunski}
\label{S:strafunski}

The two simple combinators \adhoc\ and \hfoldr\ are sufficient to
obtain the full power of strategic programming. We will demonstrate
this by the reconstruction of essential parts of the strategic
programming library of \strafunski. In fact, all previously published
examples of functional strategy
combinators~\cite{LV02-PADL,Laemmel02-TGR,LV03-PADL} can be
reconstructed with the identified primitives.

\medskip

\noindent
Normally we distinguish two broad categories of strategies, namely
\emph{type-preserving} vs.\ \emph{type-unifying} ones. In case of the
former, input and output term are of the same type. In case of the
latter, the output type is fixed regardless of the input type. Another
dimension of categorisation arises from the issue of possibly
`effectful' traversal. In previous work, all our strategy combinators
adhered to monadic style to be prepared for effects such as
partiality, environment propagation, and I/O. If no such effect is
present, the trivial identity monad can be used to `recover' from
monadic style. In the present paper, we explicitly distinguish
\emph{monadic} and \emph{non-monadic} strategies. This allows a
strategic programmer to resort to the simpler non-monadic types
whenever this is sufficient. The variation in the two aforementioned
dimensions~---~type-preservation vs.\ type-unification, and
non-monadic vs.\ monadic strategies~---~can be captured by the
following Haskell type synonyms:

\input{snip/tptu.math}

\noindent
Note that these strategy types are just instances of the more abstract
type $\mathit{Strategic}$ which we defined in the previous section. This
can be demonstrated as follows:

\noindent
\parbox{.5\textwidth}{%
\input{snip/yatptu.math}
}\ \ \ \ \parbox{.4\textwidth}{%
\input{snip/ic.math}
}

\noindent
The synonyms $\mathit{I}$, $\mathit{C}$, $\mathit{MI}$, and
$\mathit{MC}$ are Haskell implementations of non-monadic and monadic
versions of the identity and constant type constructors. The four
broad categories of strategy types are reconstructed by passing one of
these type constructors to $\mathit{Strategic}$. We assume specialised
versions of \adhoc\ and \hfoldr\ for each of these categories. We
postfix the combinators by the corresponding category. These
specialised combinators shall be used in strategic code if a specific
category is intended.\footnote{So we should have used
$\mathit{hfoldrMTU}$ instead of \hfoldr\ in the definition of
$\mathit{oneMTU}$ in Sec.~\ref{S:just}.} The specialised
combinators for the category $\mathit{TP}$, for example, are declared
as follows:

\input{snip/specialize0.math}

\noindent
These specialisations allow us to maintain ``strategy application =
function application'' in absence of type-level lambdas in
Haskell. Recall that the types of \adhoc\ and \hfoldr\ involve a
parameter $\cod$ for co-domain construction. Any instantiation of the
combinator types had to use \emph{datatypes} or \emph{newtypes} for
$\cod$ as opposed to type synonyms. This is no problem for the
category $\mathit{MTP}$ where a monad instantiates $\cod$ but the
other categories were defined above via type synonyms for identity and
constant type construction. We can easily match up the types of
\adhoc\ and \hfoldr\ with the simple types favoured by the
programmer. The trick is to wrap and unwrap extra `coaching'
constructors inside the definitions of the specialisations. We
illustrate this technique for $\mathit{adhocTP}$:

\input{snip/adhocTP.math}

\noindent
At this point, we have defined all helper types and specialised
combinators so that strategic programming can commence. We can define
numerous one-layer traversal combinators in terms of \hfoldr, e.g.:

\medskip

\input{snip/one-step.math}

\noindent
These examples reconstruct combinators as introduced
in~\cite{LV02-PADL}. The first combinator, $\mathit{oneMTU}$, was
already defined in Sec.~\ref{S:just}; here, we also provide its type,
and we use the specialised combinator $\mathit{hfoldrMTU}$ to reflect
the kind of strategy at hand. There is a class constraint
$\mathit{MonadPlus}\,m$ because $\mathit{oneMTU}$ finds the suitable
child via try and failure. The second combinator, $\mathit{allTU}$, uses
monoid operations to combine the results of applying the type-unifying
argument strategy to all children. The third combinator,
$\mathit{allTP}$, is a variation on the folklore list $\mathit{map}$.
The type-preserving argument strategy $s$ is mapped over the children
of a term. In the case of a non-empty constructor application, $s$ is
applied to the head $h$, and then the result is passed to the
recursively processed tail. The empty constructor application is
simply preserved via the identity function $\mathit{id}$. The last
combinator, $\mathit{allMTP}$, is a monadic variation on
$\mathit{allTP}$. It uses monadic bind ($\mathit{\bind}$) to sequence
the applications of the type-preserving argument strategy to all
children, and reconstructs the term with the processed children.

\medskip

\noindent
Recursive traversal combinators can now be fabricated. The following
portfolio provides \emph{traversal schemes} that are parameterised by
strategies for node-processing:

\medskip

\input{snip/recursive.math}

\noindent
The first two combinators model non-monadic, type-preserving, and full
traversal, i.e., they visit every node in the input term and the
result type coincides precisely with the type of the input type.  The
monadic, type-preserving combinator $\mathit{stop\_tdMTP}$ performs a
\emph{partial} traversal, since it does not descend below nodes where
its argument strategy is applied successfully. The type-unifying
combinator $\mathit{collect}$ performs a full traversal while
intermediate list results are concatenated. The monadic, type-unifying
combinator $\mathit{select}$ performs a partial traversal following
the same scheme as $\mathit{query}$ from the sample section. However,
the argument for recognition and extraction is not a function on a
certain term type, but a strategy. This generality is appropriate
whenever `relevant' subterms can be of different types. All these
combinators are reconstructions of combinators that are available in
the strategy library distributed with \strafunski.

%%%%%%%%%%%%%%%%%%%%%%%%%%%%%%%%%%%%%%%%%%%%%%%%%%%%%%%%%%%%%%%%%%%%%%%%%%%%%%

\section{Concluding remarks}
\label{S:concl}

We have realized a compact amalgamation of term rewriting strategies
and functional programming with emphasis on traversal
strategies. Functional strategic programming features first-class
generic functions that traverse terms of any type while mixing uniform
and type-specific behaviour. Our reconstruction of functional
strategies is based on just two combinators. Our first combinator is
\adhoc\ for type-based function dispatch. Our second combinator is
\hfoldr\ for folding over constructor applications. We have given
concise definitions of the these two basic combinators. We have
demonstrated how they are used to define one-layer traversal
combinators and recursive traversal combinators as used in strategic
programming. The abilities to traverse terms and to mix uniform and
type-specific behaviour provide, in our experience, the key to
practical application of functional programming to program analysis
and transformation problems.

\medskip

\noindent
We have discussed implementational models of functional strategies
without commitment to a specific functional language. We have, for
example, argued that adding \dimit{built-in} support for \adhoc\ and
\hfoldr\ to functional language implementations requires only modest
modifications, if second-order polymorphism is available. More
specifically, using a reference model which requires generative
support \dimit{per-type}, we have shown how the Haskell language needs
to be extended to include our two combinators. Fully operational
support for functional strategic programming is available in the form
of the \strafunski\ bundle for generic programming and language
processing in Haskell. The bundle can be configured to use one out of
several alternative models. Generative support is based on the DrIFT
preprocessing technology for Haskell. \strafunski\ has been applied
for Java refactoring, Cobol reverse engineering, grammar engineering,
Haskell program analysis and transformation, XML document
transformation, and others.

\medskip

\noindent
Thus, our approach is lightweight, highly expressive, well-founded,
and has already proven its value in important application domains.
Language users take advantage of the generality and simplicity
provided by our combinator style of generic programming. Language
implementors take advantage of the fact that strategic programming
does not require any new language constructs, but only two simple
combinators which are easily defined \dimit{per-type} or as
\dimit{built-in}s.  There is no need for compile-time specialisation,
and separate compilation is easily maintained.

%%%%%%%%%%%%%%%%%%%%%%%%%%%%%%%%%%%%%%%%%%%%%%%%%%%%%%%%%%%%%%%%%%%%%%%%%%%%%%

\paragraph*{Acknowledgement} We are grateful for discussions with
David Basin, Robert Ennals, Kevin Hammond, Ralf Hinze, Barry Jay,
Johan Jeuring, Claude Kirchner, Paul Klint, Jan Kort, Andres L{\"o}h,
Simon Peyton Jones, Claus Reinke, Simon Thompson, Phil Trinder, Jurgen
Vinju, and Stephanie Weirich.

%%%%%%%%%%%%%%%%%%%%%%%%%%%%%%%%%%%%%%%%%%%%%%%%%%%%%%%%%%%%%%%%%%%%%%%%%%%%%%

\bibliographystyle{abbrv}
\bibliography{paper}

%%%%%%%%%%%%%%%%%%%%%%%%%%%%%%%%%%%%%%%%%%%%%%%%%%%%%%%%%%%%%%%%%%%%%%%%%%%%%%

\end{document}

%% file: snip/recursive.math
{\setlength{\lwidth}{0.5\lwidth}
\noindent
\hscaption{Full traversal of all nodes in top-down manner}\begin{tabbing}
\qquad\=\hspace{\lwidth}\=\hspace{\cwidth}\=\+\kill
$\Varid{full\char95 tdTP}$ \> \makebox[\cwidth]{$\mathbin{::}$} \> ${\Conid{TP}\to \Conid{TP}}$\\
$\Varid{full\char95 tdTP}\;\Varid{s}\;\Varid{x}$ \> \makebox[\cwidth]{$\mathrel{=}$} \> ${\Varid{allTP}\;(\Varid{full\char95 tdTP}\;\Varid{s})\;(\Varid{s}\;\Varid{x})}$
\end{tabbing}\hscaption{Full traversal of all nodes in bottom-up manner}\begin{tabbing}
\qquad\=\hspace{\lwidth}\=\hspace{\cwidth}\=\+\kill
$\Varid{full\char95 buTP}$ \> \makebox[\cwidth]{$\mathbin{::}$} \> ${\Conid{TP}\to \Conid{TP}}$\\
$\Varid{full\char95 buTP}\;\Varid{s}\;\Varid{x}$ \> \makebox[\cwidth]{$\mathrel{=}$} \> ${\Varid{s}\;(\Varid{allTP}\;(\Varid{full\char95 buTP}\;\Varid{s})\;\Varid{x})}$
\end{tabbing}\hscaption{Top-down traversal with cut after success}\begin{tabbing}
\qquad\=\hspace{\lwidth}\=\hspace{\cwidth}\=\+\kill
$\Varid{stop\char95 tdMTP}$ \> \makebox[\cwidth]{$\mathbin{::}$} \> ${\forall{m}.\ \Conid{MonadPlus}\;\Varid{m}\Rightarrow \Conid{MTP}\;\Varid{m}\to \Conid{MTP}\;\Varid{m}}$\\
$\Varid{stop\char95 tdMTP}\;\Varid{s}\;\Varid{x}$ \> \makebox[\cwidth]{$\mathrel{=}$} \> ${\Varid{s}\;\Varid{x}\mathbin{`\Varid{mplus}`}\Varid{allMTP}\;(\Varid{stop\char95 tdMTP}\;\Varid{s})\;\Varid{x}}$
\end{tabbing}\hscaption{Accumulating a list query all-over the place}\begin{tabbing}
\qquad\=\hspace{\lwidth}\=\hspace{\cwidth}\=\+\kill
$\Varid{collect}$ \> \makebox[\cwidth]{$\mathbin{::}$} \> ${\forall{u}.\ \Conid{TU}\;[\mskip1.5mu \Varid{u}\mskip1.5mu]\to \Conid{TU}\;[\mskip1.5mu \Varid{u}\mskip1.5mu]}$\\
$\Varid{collect}\;\Varid{s}\;\Varid{x}$ \> \makebox[\cwidth]{$\mathrel{=}$} \> ${\Varid{s}\;\Varid{x}\plus \Varid{allTU}\;(\Varid{collect}\;\Varid{s})\;\Varid{x}}$
\end{tabbing}\hscaption{Selection from the first node that admits success}\begin{tabbing}
\qquad\=\hspace{\lwidth}\=\hspace{\cwidth}\=\+\kill
$\Varid{select}$ \> \makebox[\cwidth]{$\mathbin{::}$} \> ${\forall{u}~{m}.\ \Conid{MonadPlus}\;\Varid{m}\Rightarrow \Conid{MTU}\;\Varid{u}\;\Varid{m}\to \Conid{MTU}\;\Varid{u}\;\Varid{m}}$\\
$\Varid{select}\;\Varid{s}\;\Varid{x}$ \> \makebox[\cwidth]{$\mathrel{=}$} \> ${\Varid{s}\;\Varid{x}\mathbin{`\Varid{mplus}`}\Varid{oneMTU}\;(\Varid{select}\;\Varid{s})\;\Varid{x}}$
\end{tabbing}}